\documentstyle[11pt,epsfig]{article}
\newcounter{subequation}[equation]

\makeatletter 
 
\def\thesubequation{\theequation\@alph\c@subequation}
\def\@subeqnnum{{\rm (\thesubequation)}}
\def\slabel#1{\@bsphack\if@filesw {\let\thepage\relax
   \xdef\@gtempa{\write\@auxout{\string
      \newlabel{#1}{{\thesubequation}{\thepage}}}}}\@gtempa
   \if@nobreak \ifvmode\nobreak\fi\fi\fi\@esphack}
\def\subeqnarray{\stepcounter{equation}
\let\@currentlabel=\theequation\global\c@subequation\@ne
\global\@eqnswtrue
\global\@eqcnt\z@\tabskip\@centering\let\\=\@subeqncr
$$\halign to \displaywidth\bgroup\@eqnsel\hskip\@centering
  $\displaystyle\tabskip\z@{##}$&\global\@eqcnt\@ne
  \hskip 2\arraycolsep \hfil${##}$\hfil
  &\global\@eqcnt\tw@ \hskip 2\arraycolsep
  $\displaystyle\tabskip\z@{##}$\hfil
   \tabskip\@centering&\llap{##}\tabskip\z@\cr}
\def\endsubeqnarray{\@@subeqncr\egroup
                     $$\global\@ignoretrue}
\def\@subeqncr{{\ifnum0=`}\fi\@ifstar{\global\@eqpen\@M
    \@ysubeqncr}{\global\@eqpen\interdisplaylinepenalty \@ysubeqncr}}
\def\@ysubeqncr{\@ifnextchar [{\@xsubeqncr}{\@xsubeqncr[\z@]}}
\def\@xsubeqncr[#1]{\ifnum0=`{\fi}\@@subeqncr
   \noalign{\penalty\@eqpen\vskip\jot\vskip #1\relax}}
\def\@@subeqncr{\let\@tempa\relax
    \ifcase\@eqcnt \def\@tempa{& & &}\or \def\@tempa{& &}
      \else \def\@tempa{&}\fi
     \@tempa \if@eqnsw\@subeqnnum\refstepcounter{subequation}\fi
     \global\@eqnswtrue\global\@eqcnt\z@\cr}
\let\@ssubeqncr=\@subeqncr
\@namedef{subeqnarray*}{\def\@subeqncr{\nonumber\@ssubeqncr}\subeqnarray}
\@namedef{endsubeqnarray*}{\global\advance\c@equation\m@ne%
                           \nonumber\endsubeqnarray}

\makeatletter
\@addtoreset{equation}{section}
\makeatother
\renewcommand{\theequation}{\thesection.\arabic{equation}}

\def\dalemb#1#2{{\vbox{\hrule height .#2pt
        \hbox{\vrule width.#2pt height#1pt \kern#1pt
                \vrule width.#2pt}
        \hrule height.#2pt}}}

    \let\e=\epsilon
  \let\q=\theta  
  \let\n=\nu

\def\nn{\nonumber} \def\bd{\begin{document}} \def\ed{\end{document}}
\def\ds{\documentstyle} \let\fr=\frac \let\bl=\bigl \let\br=\bigr
\let\Br=\Bigr \let\Bl=\Bigl 
\let\bm=\bibitem
\let\na=\nabla
\let\pa=\partial \let\ov=\overline
\def\ie{{\it i.e.\ }} 
\newcommand{\be}{\begin{equation}} 
\newcommand{\ee}{\end{equation}} 
\def\ba{\begin{array}}
\def\ea{\end{array}}
\def\ft#1#2{{\textstyle{{\scriptstyle #1}\over {\scriptstyle #2}}}}
\def\fft#1#2{{#1 \over #2}}
\def\del{\partial}
\def\sst#1{{\scriptscriptstyle #1}}
\def\oneone{\rlap 1\mkern4mu{\rm l}}
\def\e7{E_{7(+7)}}
\def\td{\tilde}
\def\wtd{\widetilde}
\def\im{{\rm i}}
\def\bog{Bogomol'nyi\ }
\def\q{{\tilde q}}
\def\hast{{\hat\ast}}
\def\0{{\sst{(0)}}}
\def\1{{\sst{(1)}}}
\def\2{{\sst{(2)}}}
\def\3{{\sst{(3)}}}
\def\4{{\sst{(4)}}}
\def\5{{\sst{(5)}}}
\def\6{{\sst{(6)}}}
\def\7{{\sst{(7)}}}
\def\8{{\sst{(8)}}}
\def\n{{\sst{(n)}}}
\def\oo{{\"o}}
\def\hA{\hat{\cal A}}
\def\ns{{\sst {\rm NS}}}
\def\rr{{\sst {\rm RR}}}
\def\tH{{\widetilde H}}
\def\tB{{\widetilde B}}
\def\cA{{\cal A}}
\def\cF{{\cal F}}
\def\tF{{\wtd F}}
\def\Z{\rlap{\sf Z}\mkern3mu{\sf Z}}
\def\ep{{\epsilon}}
\def\IIA{{\rm IIA}}
\def\IIB{{\rm IIB}}
\def\ads{{\rm AdS}}
\def\R{\rlap{\rm I}\mkern3mu{\rm R}}
\def\mapright#1{\smash{\mathop{-\!\!\!-\!\!\!-\!\!\!-\!\!\!-\!\!\!
             \longrightarrow}\limits^{#1}}}

\def\Ei{{\hbox{Ei}}}
\def\Ci{{\hbox{Ci}}}
\def\Si{{\hbox{Si}}}

\newcommand{\ho}[1]{$\, ^{#1}$}
\newcommand{\hoch}[1]{$\, ^{#1}$}
\newcommand{\bea}{\begin{eqnarray}} 
\newcommand{\eea}{\end{eqnarray}} 
\newcommand{\ra}{\rightarrow}
\newcommand{\lra}{\longrightarrow}
\newcommand{\Lra}{\Leftrightarrow}
\newcommand{\aap}{\alpha^\prime}
\newcommand{\bp}{\tilde \beta^\prime}
\newcommand{\tr}{{\rm tr} }
\newcommand{\Tr}{{\rm Tr} } 
\newcommand{\NP}{Nucl. Phys. }
\newcommand{\upenn}{\it Dept. of Phys. and Astro., 
University of Pennsylvania,
Philadelphia, PA 19104}

\newcommand{\auth}{J. F. V\'azquez-Poritz}

\begin{document}
\begin{flushright}
UPR/898-T \\
July 2000\\
\hfill{\bf hep-th/0007202}\\
\end{flushright}


\begin{center}

{\large {\bf  Absorption by Nonextremal D3-branes}
}

\vspace{20pt}

\auth

\vspace{10pt}
{\hoch{\dagger}\upenn}

\vspace{30pt}

\underline{ABSTRACT}
\end{center}

We calculate the absorption probabilities for a class of massless fields 
whose linear perturbations leave the near-extremal D3-brane background
metric unperturbed. It has previously been found that, for extremal
D3-branes, these fields share the same absorption probability as that of 
the dilaton-axion. We find that these absorption probabilities diverge
from each other as we move away from extremality. The form of the
corresponding effective Schr\"odinger potentials leads us to conjecture that 
the absorption of various fields by nonextremal D3-branes depends on the
polarization of angular momentum.

{\vfill\leftline{}\vfill
\vskip 10pt \footnoterule {\footnotesize \hoch{1} Research supported
in part by DOE grant DOE-FG02-95ER40893}

\pagebreak
\setcounter{page}{1}


\section{Introduction}

Over the past few years, there has been much work done on the
scattering of fields due the curved backgrounds of p-brane configurations
of M theory and string theory \cite{raams,gkt,kleb,ghkk,emp,taylor,gubser,
gubserhash,clpt,clpt2,lee}. Most of this research has focused on the
scattering of minimally-coupled massless scalar fields, though there has
been some work on fermions \cite{hoso}, two-form fields \cite{raja} and 
massive scalars \cite{mass}. The absorption probabilities of various linear 
field perturbations for which the ten-dimensional metric is left
unperturbed have been calculated for extremal D3-branes \cite{mathur}. It
has been shown, through the analysis of wave equations in Schr\"odinger
form, that certain half-integer and integer spin massless modes of the
extremal D3-brane have identical absorption probabilities \cite{poritz},
which implies that such fields couple on the dual field theory to
operators forming supermultiplets of strongly coupled gauge theory. In an
effort to enhance our understanding of non-BPS states, the present paper
continues this study for non-extremal D3-branes, which have already
been probed in the near-extremal case by minimally-coupled massless
scalars \cite{siopsis,siopsis2}. 

Our discussion is organized as follows. In section II, we present the wave
equations for various linear field perturbations in a non-extremal
D3-brane background, for which the ten-dimensional metric is left
unperturbed. In section III, we calculate the low-energy absorption
probabilities for these field perturbations in the background of
near-extremal D3-branes, using a method that was initially applied to a
minimally-coupled scalar probe \cite{siopsis,siopsis2}. In section IV, we
provide concluding remarks.

\section{Effective potentials}

The D3-brane of type IIB supergravity is given by
\begin{eqnarray}
ds^2_{10}&=&H^{-1/2} (-f dt^2+ dx_1^2 +dx_2^2+dx_3^2) +
H^{1/2} (f^{-1} dr^2 + r^2 d\Omega_{5}^2), \\ \nonumber
G_{(5)}&=&d^4 x \wedge dH^{-1}+\ast (d^4 x \wedge
dH^{-1}). \label{d3metric}
\end{eqnarray}
where $H=1+\frac{R^4}{r^4}$ and $f=1 - \frac{r_o^4}{r^4}$. $R$ specifies
the D3-brane charge and $r_o$ is the non-extremality parameter. The field
equations are
\begin{eqnarray}
R_{\mu \nu} &=& -\frac{1}{6}F_{\mu \rho \sigma \tau \kappa}F_{\nu} ^{\rho
\sigma \tau \kappa}, \\ \nonumber
F_{\mu \nu \rho \sigma \tau}&=&\frac{1}{5!}\epsilon_{\mu \nu \rho \sigma
\tau \mu^{'} \nu^{'} \rho^{'} \sigma^{'} \tau^{'}}F^{\mu^{'} \nu^{'}
\rho^{'} \sigma^{'} \tau^{'}}, \\ \nonumber
D^{\mu} \partial_{[\mu}A_{\nu \rho ]} &=& -\frac{2i}{3} \dot{F}_{\nu \rho
\sigma \tau \kappa} D^{\sigma} A^{\tau \kappa}, \\ \nonumber
D^{\mu} \partial_{\mu} B &=& 0,
\end{eqnarray}
where a dot above a symbol for a field denotes its background value.
We study the wave equations for linear perturbations that leave the
ten-dimensional background metric unperturbed. Deriving the radial wave
equations in the background of non-extremal D3-branes is a
straightforward generalization of what is done in \cite{mathur} for the
extremal case.

\subsection{Dilaton-axion}

The dilaton and axion are decoupled from the D3-brane in type IIB theory,
satisfying the minimally-coupled scalar wave equation
\be
\frac{1}{\sqrt{-g}} \partial_{\mu} \sqrt{-g} g^{\mu \nu} \partial_{\nu}
\phi=0. \label{dilaton}
\ee
Thus, the radial wave equation of a dilaton-axion in the spacetime of a
non-extremal D3-brane is given by
\be
\Big( \frac{1}{r^5} \frac{\partial}{\partial r} f r^5 
\frac{\partial}{\partial r} + \omega^2 \frac{H}{f} - 
\frac{\ell (\ell + 4)}{r^2} \Big) \phi = 0, \label{D3eqn}
\ee
where $\ell=0,1,..$

We shall express the wave equation in Schr\"odinger form, and study the
characteristics of the Schr\"odinger effective potential.  By the
substitution
\be
\phi = r^{-5/2} f^{-1/2} \psi,
\ee
we render (\ref{D3eqn}) in the Schr\"odinger form
\be
\big(\frac{\partial^2}{\partial r^2}-V_{\rm eff} \big)\psi = 0,
\label{eqdil}
\ee
where 
\be
V_{\rm eff}(\ell) =-\frac{\omega^2}{f^2}H+\frac{(\ell+3/2)(\ell+5/2)}{f
r^2}+\frac{(f-1)(-f+16)}{4f^2r^2}. \label{V}
\ee
Factors that are shared by the incident and outgoing parts of the wave
function cancel out when calculating the absorption probability. Thus, the
absorption probability of $\phi$ and $\psi$ are the same. 

Technically, $V_{\rm eff}$ cannot be interpreted as an effective potential,
since it is dependent on the particle's incoming energy. However, this
is of no consequence for our analysis of the form of the wave equations
for various particles.

\subsection{Scalar from the two-form}

For the free indices of the two-form taken to lie along the $S^5$, the
radial wave equation is 
\be
\Big( \frac{H}{r} \frac{\partial}{\partial r} \frac{f r}{H} 
\frac{\partial}{\partial r} + \frac{H}{f} \omega^2 -
\frac{(\ell+2)^2}{r^2} \mp \frac{4R^4}{H r^6} (\ell+2) \Big) a_{(\alpha
\beta)} = 0, \label{scalar}
\ee
where $\ell=1,2,..$
The sign $\pm$ corresponds to the sign in the spherical harmonic
involved in the partial wave expansion.

By the substitution
\be
a_{(\alpha \beta)} = \big( \frac{H}{f r} \big)^{1/2} \psi_{(\alpha
\beta)},
\ee
we render (\ref{scalar}) into the Schr\"odinger form:
\begin{eqnarray}
V_{\rm eff}(\ell) = -\frac{\omega^2 H}{f^2}+\frac{(\ell+3/2)(\ell+5/2)}{f
r^2}+\frac{(f-1)(15f+16)}{4r^2 f^2}+ \nn\\ 
\frac{(H-1)[-f^2(15H+49)+f (\pm 16(\ell+2)H+30H+2)+H-1]}{4r^2 f^2 H^2}.
\label{Vscalar}
\end{eqnarray}
\subsection{Vector from the two-form}

We now consider one free index of the two-form along $S^5$ and one
free index in the remaining 5 directions. For the tangential components of
the vector, the radial wave equation is
\be
\Big( \frac{1}{r^3} \frac{\partial}{\partial r} r^3 f 
\frac{\partial}{\partial r} + \frac{\omega^2 H}{f} - 
\frac{(\ell+1)(\ell+3)}{r^2} \Big) a_1 = 0, \label{tang}
\ee
where $\ell=1,2,..$

By the substitution
\be
a_1  = (r^3 f)^{-1/2} \psi,
\ee
we render (\ref{tang}) in the Schr\"odinger form with
\be
V_{\rm eff}(\ell) = -\frac{\omega^2 H}{f^2}+\frac{\ell+3/2)(\ell+5/2)}{f
r^2} +\frac{(f-1)(3f+16)}{4f^2r^2}.
\ee
The radial and time-like components of the vector can be determined from each
other by
\be
\frac{\partial}{\partial r} a_o = \frac{1}{i \omega} \big[
\omega^2-\frac{(\ell+1)(\ell+3)}{r^2 H}f \big] a_r \label{a}
\ee
and
\be
\frac{1}{r^3}\frac{\partial}{\partial r} \big[ r^3 \big(
\frac{\partial}{\partial r} a_0+i \omega a_r \big) \big]-\frac{(\ell+1)
(\ell+3)}{fr^2} a_0=0, \label{aa}
\ee
where $\ell=1,2,..$

(\ref{a}) and (\ref{aa}) can be decoupled. The wave equation for
$a_r$ is
\be
\Big( H \frac{\partial}{\partial r} \frac{f}{r} 
\frac{\partial}{\partial r} \frac{r f}{H} + \omega^2 H - 
\frac{(\ell+1)(\ell+3)}{r^2}f \Big) a_r = 0. \label{ar}
\ee
By the substitution
\be
a_r = (r f^3)^{-1/2} H \psi, \label{aar}
\ee
we render (\ref{ar}) into the Schr\"odinger form with
\be
V_{\rm eff}(\ell) = -\frac{\omega^2 H}{f^2}+\frac{\ell+3/2)(\ell+5/2)}{f
r^2} +\frac{(f-1)(35f+16)}{4f^2r^2}.
\ee
The wave equation for $a_0$ is
\be
\Big( \frac{f}{r} \frac{\partial}{\partial r}
\frac{r^3 f}{\omega^2 r^2 H-(\ell+1)(\ell+3)f} \frac{\partial}{\partial
r} +1 \Big) a_0=0. \label{a0}
\ee
By the substitution
\be
a_0=(\frac{r^3 f}{\omega^2 r^2 H-(\ell+1)(\ell+3)f})^{-1/2} \psi,
\label{ao}
\ee
(\ref{a0}) can be rendered into the Schr\"odinger form. However, the
corresponding potential is singular. Instead, we determine $a_0$ directly
from $a_r$ via
\be
a_0=\frac{i f}{\omega r} \frac{\partial}{\partial r} \Big( \frac{rf}{H}
a_r \Big). \label{aoar}
\ee

\subsection{Antisymmetric tensor from 4-form}

For two free indices of the 4-form along $S^5$ and two
free indices in the remaining 5 directions, the coupled radial wave
equations for the components of the antisymmetric tensor derived from the
4-form are
\be
b_{3r}=\mp \frac{\omega rH}{\ell+2} \frac{1}{f} b_{12},
\ee
\be
\frac{\partial}{\partial r}b_{03}-i \omega b_{3r}=\pm \frac{i}{r} (\ell+1)
b_{12}
\ee
and
\be
\frac{\partial}{\partial r} b_{12}=\mp \frac{i}{r f} (\ell+2)b_{03},
\ee
where $\ell=1,2,..$
Eliminating $b_{03}$ and $b_{3r}$ we get
\be
\Big( \frac{1}{r} \frac{\partial}{\partial r} r f \frac{\partial}{\partial
r} +\frac{\omega^2 H}{f}-\frac{(\ell+2)^2}{r^2} \Big) b_{12}=0,
\label{4-form}
\ee
By the substitution
\be
b_{12}=(r f)^{-1/2} \psi,
\ee
we render (\ref{4-form}) into Schr\"odinger form with
\be
V_{\rm eff}(\ell) = -\frac{\omega^2 H}{f^2}+\frac{(\ell+3/2)(\ell+5/2)}{f
r^2} +\frac{(f-1)(15f+16)}{4f^2r^2}.
\ee

\subsection{Two-form from the antisymmetric tensor}

The equations for two-form perturbations polarized along the
D3-brane are coupled. For s-wave perturbations, they can be
decoupled \cite{raja,mathur}, and the radial wave equation is
\be
\Big( \frac{1}{r^5 H} \frac{\partial}{\partial r} r^5 f H 
\frac{\partial}{\partial r} + \frac{\omega^2 H}{f} - 
\frac{16 R^8}{r^{10} H^2} \Big) \phi = 0, \label{2-form}
\ee
where $\ell=0,1,..$

By the substitution
\be
\phi = (r^5 f H)^{-1/2} \psi,
\ee
we render (\ref{2-form}) into Schr\"odinger form with
\be
V_{\rm eff}(\ell) = -\frac{\omega^2 H}{f^2}+\frac{15}{4f r^2}
+\frac{12(H-1)^2}{r^2 f H^2}+\frac{(f-1)(15f+16)}{4r^2 f^2}. \label{2form}
\ee

\subsection{General form of effective potential}

For an extremal D3-brane, the effective potentials for the dilaton-axion,
vector from the two-form and the antisymmetric tensor from the 4-form all
reduce to
\be
V_{\rm eff}(\ell) = -\omega^2 H+\frac{(\ell+3/2)(\ell+5/2)}{r^2}.
\ee
Thus, the above fields have identical absorption probabilities. The
effective potentials for the scalar from the two-form and the two-form
from the antisymmetric tensor include additional terms, which
merely have the effect of changing the partial wave number by
$\ell \rightarrow \ell \pm 1$ \cite{poritz}. 

For a non-extremal D3-brane, the effective potentials for the dilaton-axion,
vector from the two-form and the antisymmetric tensor from the 4-form are
of the form
\be
V_{\rm eff}(\ell) = -\frac{\omega^2 H}{f^2}
+\frac{(\ell+3/2)(\ell+5/2)}{f r^2}+
\frac{\big(f-1\big)\big((a-1)(a+1) f+16\big)}{4f^2 r^2}, \label{general}
\ee
where $a=0$ for the dilaton-axion, $a=2$ for the tangential components of
the vector from the two-form, $a=4$ for the antisymmetric tensor from the
4-form, and $a=6$ for the radial component of the vector from the
two-form. For the scalar from the two-form and the two-form from the
antisymmetric tensor, the effective potential has additional terms that
are proportional to the charge of the D3-brane. However, for a chargeless
D3-brane, the scalar and two-form fields fit into the above scheme with
$a=4$. It is interesting to note the change of grouping of absorption
probabilities between the cases of extremal and chargeless D3-branes. 
The effective potential for the time-component of the vector from
the two-form does not appear to fit into this scheme, though certain
potential terms may merely have the effect of changing the partial wave number.

We conjecture that $a$ is a parameter that depends on the polarization of
the angular momentum. Note that this parameter only plays a role in
absorption away from extremality.

\section{Low-energy absorption probabilities for a near-extremal
D3-brane}

\subsection{Dilaton-axion, vector from two-form, and antisymmetric tensor 
from 4-form}

We will solve the wave equations in the approximation that
\be
r_o \ll R \ll 1/\omega.
\ee
We work in the limit in which the frequency is large compared to the
temperature, which means that $r_o \leq R^2 \omega$. We render the wave
function in a form that will be convenient for isolating the singularity
at $r=r_o$:
\be
\psi=(r^5 f)^{+1/2} \tilde{\phi}. \label{tilde}
\ee
Substituting (\ref{tilde}) into (\ref{eqdil}) together with
(\ref{general}) yields
\be
\Big( \frac{1}{r^5} \partial_r fr^5 \partial_r +\omega^2 \frac{H}{f}
-\frac{\ell(\ell+4)}{r^2}+\frac{a^2}{4r^2} (1-f) \Big) 
\tilde{\phi}=0. \label{general2}
\ee
Note that $\phi=F(r) \tilde{\phi}$, where $F(r)$ is a nonsingular function
of $r$ for $r>r_o$.

We will now use the same approximation and procedure as that by Siopsis in
the case of a minimally-coupled scalar \cite{siopsis,siopsis2}. We use three 
matching regions. In the outer region defined by $r \gg R^2 \omega$,
(\ref{general2}) becomes 
\be
\Big( r^2 \partial_r^2 +5r\partial_r +\omega^2
r^2-\ell(\ell+4) \Big) \tilde{\phi}=0. \label{outerV}
\ee
The solution is
\be
\tilde{\phi}=\frac{1}{r^2}J_{(\ell+2)}(\omega r). \label{outersol}
\ee
In the intermediate region defined by $r-r_o \gg r_o$ and $r \ll R$,
expressed in terms of the dimensionless quantity $z \equiv (r-r_o)/r_o$,
\be
\Big( \frac{1}{z^3} \partial_z z^5 \partial_z +\frac{16 
\kappa^2}{z^2}-\ell(\ell+4) \Big) \tilde{\phi}=0,
\ee
where
\be
\kappa=\frac{\omega r_o}{4} \big( 1+\frac{R^4}{r_o^4} \big) ^{1/2} \approx
\frac{R^2 \omega}{4 r_o}=\frac{\omega}{4 \pi T_H},
\ee
and $T_H=\frac{r_o}{\pi R^2}$ is the Hawking temperature. The corresponding 
wave function solution is
\be
\tilde{\phi}= \frac{A}{r_o^2 z^2} H_{\ell+2}^{(1)} \big (\frac{4\kappa}
{z}\big), \label{intermsol}
\ee
where we consider the purely incoming solution. Matching asympototic form
of the intermediate solution  (\ref{intermsol}) for large $z$ with the
asymptotic form of the outer solution (\ref{outersol}) for small $\omega r$ 
yields the amplitude
\be
A=\frac{i \pi}{(\ell+2)!(\ell+1)!} \big( \frac{\omega R}{2}\big)
^{2\ell+4}. \label{A}
\ee
For $z \ll \kappa$, we expand the intermediate wave function solution
(\ref{intermsol}):
\be
\tilde{\phi}= \frac{(-i)^{\ell+5/2} A}{r_o^2 \sqrt{2\pi \kappa}} z^{-3/2}
e^{4i\kappa /z} \Big(
1+\frac{i(\ell+3/2)(\ell+5/2)z}{8\kappa}+O(\kappa^{-2})\Big). \label{small}
\ee
The inner region is defined by $r-r_o \ll R^2 \omega$. The inner and 
intermediate regions overlap, since $r_o \ll R^2 \omega$. There is a
singularity in the wave function at $r=r_o$, which can be isolated by
taking the wave function to be \cite{siopsis,siopsis2}
\be
\tilde{\phi}= f^{i \kappa} \varphi. \label{innersol}
\ee
Isolating this singularity in the wavefunction enables us to calculate the
dominant term in the near-extremal absorption probability. 
For the inner region, we substitute (\ref{innersol}) into 
(\ref{general2}) and express the result in terms of the dimensionless
parameter $x\equiv r/r_o$:
\be
x^2 f(x) \partial_x^2 \varphi+x\big[
5-(1-8i\kappa) x^{-4} \big] \partial_x \varphi +\Big(16 \kappa^2 h(x)-
\ell(\ell+4) +\frac{a^2}{4}(1-f(x)) \Big) \varphi=0, \label{innerequation}
\ee
where
\be
f(x)=1-x^{-4} \label{f}
\ee
and
\be
h(x) \equiv \frac{1+x^2+x^4}{(1+x^2)x^4}.\label{h}
\ee
Since $x \ll \kappa$, we shall expand the inner wave function solution
in $\kappa^{-1}$:
\be
\varphi = A B e^{i\kappa \alpha (x)} \beta(x) \big( 1+\frac{i}{\kappa}
\gamma(x) + O(\kappa^{-2}) \big). \label{solform}
\ee
We will solve for B such that $\varphi(1)=B$, which implies that
$\alpha(1)=\gamma(1)=0$ and $\beta(1)=1$. We solve for the functions
$\alpha$,$\beta$ and $\gamma$ by plugging (\ref{solform}) into 
(\ref{innerequation}):
\be
\alpha (x) = -4 \int_{1}^{x} dy
\frac{1+\sqrt{1+h(y)y^4(y^4-1)}}{y(y^4-1)}.
\ee
\be
\beta (x) = \exp \Big( -\frac{1}{2} \int_{1}^{x} dy
\frac{y(y^4-1)\alpha^{''} +(5y^4-1)\alpha^{'}}{4+y(y^4-1)\alpha^{'}}
\Big).
\ee
\be
\gamma (x)=-\frac{1}{2} \int_{1}^{x} dy \frac{y^4 \ell(\ell+4)-y(y^4-1) 
\frac{\beta^{''}}{\beta}-(5y^4-1)\frac{\beta^{'}}{\beta}-\frac{a^2}{4y}}
{y(y^4-1)\alpha^{'}+4}.
\ee
We match the expressions for $\alpha$, $\beta$ and $\gamma$ in the large
$x$ limit with the asymptotic form of the intermediate solution
(\ref{small}) and solve for $B$:
\be
B=\frac{(-i)^{\ell+5/2}}{r_o^2 \sqrt{2\pi \kappa}} \big(
1-\frac{i(\ell+3/2)(\ell+5/2)}{8\kappa} \big).
\ee
The absorption probability is
\be
P=8\pi \kappa r_o^4 |A|^2 |B|^2. \label{prob}
\ee
Plugging in the amplitudes $A$ and $B$, we find that
\be
P^{near-extremal}= \Big( 1+\big( \frac{4(\ell+2)^2-1}{32 \kappa} \big)^2
\Big) P^{extremal}, \label{nearprob} 
\ee
where
\be
P^{extremal}=\frac{4 \pi^2}{\big( (\ell+1)!(\ell+2)! \big)^2} \big(
\frac{\omega R}{2} \big)^{4\ell+8}. \label{extremal}
\ee
Our result agrees with \cite{siopsis,siopsis2} for the minimally-coupled
scalar. As we have shown, for near-extremal D3-branes, the dilaton-axion,
tangential and radial components of the vector from the two-form and the
antisymmetric tensor from the 4-form have identical absorption
probabilities, as in the extremal case. As can be seen, as $\kappa
\rightarrow \infty$, we recover the result previously obtained for an 
extremal D3-brane.

\subsection{Time-like component of vector from two-form} 

Using (\ref{aar}),(\ref{aoar}) and (\ref{tilde}) we find the wave function
for the longitudinal component of the vector from the two-form directly in
terms of the radial component wave function:
\be
a_o=\frac{if}{\omega r} \partial_r (r^3 \tilde{\phi}).
\ee
We can find the wave function solutions for $a_o$ in the three
regions directly from the dilaton wave functions (\ref{outersol}),
(\ref{intermsol}) and (\ref{solform}). We must be careful to expand to
$O(\kappa^{-3})$ in the dilaton wave functions in order for $a_o$ to be of
order $O(\kappa^{-2})$. Matching $a_o$ in the three regions yields the
same absorption probability as for the dilaton given by (\ref{nearprob}).
Thus, for near-extremal D3-branes, the longitudinal component of the
vector from the two-form and dilaton-axion have identical absorption
probabilities, as in the extremal case.

\subsection{Two-form from the antisymmetric tensor}

Substituting (\ref{tilde}) into (\ref{eqdil}) together with
(\ref{2form}) yields
\be
\Big( \frac{1}{r^5} \partial_r fr^5 \partial_r +\omega^2 \frac{H}{f}
+\frac{4}{r^2}(1-f)-\frac{12(H-1)^2}{r^2 H^2} \Big) 
\tilde{\phi}=0. \label{gen2form}
\ee
We require four regions for matching. Consider $r_1$ and $r_2$ such that
$r_2 \ll R \ll r_1$. The outer region is given by $r > r_1$, where $\omega
r_1 \ll 1$. In the outer region, the wave equation and solution are given
by (\ref{outerV}) and (\ref{outersol}), respectively, with $\ell=0$.

The outer intermediate region is given by $r_1 > r > r_2$ and $\omega R^2
\ll r$, so that we can ignore the $\omega^2$ term in the wave equation:
\be
\Big( \frac{1}{r^5} \partial_r r^5 \partial_r -\frac{12(H-1)^2}{r^2 H^2} \Big) 
\tilde{\phi}=0.
\ee
The corresponding wave function solution is
\be
\tilde{\phi}= C H^{3/2} +D H^{-1/2}.
\ee
Matching the outer and outer intermediate regions yields
\be
C+D=\frac{\omega^2}{8}.
\ee
The inner intermediate region is defined by $r-r_o \gg r_o$ and $r \ll R$.
Expressed in terms of the dimensionless quantity $z \equiv (r-r_o)/r_o$,
the wave equation in this region is
\be
\big( \frac{1}{z^3} \partial_z z^5 \partial_z +\frac{16\kappa^2}{z^2}-12
\big) \tilde{\phi}.
\ee
The corresponding wave function solution is
\be
\tilde{\phi}= A \frac{1}{r_o^2 z^2} H_4^{(1)} \big( \frac{4\kappa}{z}
\big),\label{interm2form}
\ee
where we take the purely incoming solution. 
Matching the wave function across the two intermediate regions yields
\be
A=\frac{i \pi}{12} \big( \frac{\omega R}{2}\big)^{6},\ \ \ \ \ C=0.
\ee
For $z \ll \kappa$, we expand the inner intermediate wave function
solution (\ref{interm2form}):
\be
\tilde{\phi}= \frac{(-i)^{1/2} A}{r_o^2 \sqrt{2\pi \kappa}} z^{-3/2}
e^{4i\kappa /z} \big(
1+\frac{i(7/2)(9/2)z}{8\kappa}+O(\kappa^{-2})\big).\label{small2form}
\ee
For the inner region, we substitute (\ref{innersol}) into (\ref{gen2form}) 
and express the result in terms of the dimensionless parameter $x$:
\be
x^2 f(x) \partial_x^2 \varphi+x\big[
5-(1-8i\kappa) x^{-4} \big] \partial_x \varphi +\Big(16 \kappa^2 h(x)
+4\big( 1-f(x)\big)-12\Big) \varphi=0. \label{innerequation2form}
\ee
where $f(x)$ and $h(x)$ are given by (\ref{f}) and (\ref{h}),
respectively.

Following the case for the dilaton, we plug the ansatz for the inner wave
function solution, given by (\ref{solform}), into
(\ref{innerequation2form}) and solve for the amplitude $B$ by matching the
inner solution with the inner intermediate solution. The result is that
\be
B=\frac{(-i)^{1/2}}{r_o^2 \sqrt{2\pi \kappa}} \big(
1-\frac{i(7/2)(9/2)}{8\kappa} \big).
\ee
Plugging amplitudes $A$ and $B$ into (\ref{prob}), we find that the
absorption probability is
\be
P^{near-extremal}= \Big( 1+\big( \frac{63}{32 \kappa} \big)^2
\Big) P^{extremal}, 
\ee
where $P^{extremal}$ is the same as that given for the dilaton in
(\ref{extremal}) with $\ell \rightarrow \ell + 1$.
As with previous cases, as $\kappa \rightarrow \infty$, we
recover the result previously obtained for an extremal D3-brane.
We have shown that, for near-extremal D3-branes, the two-form from the
antisymmetric tensor does not have the same absorption probability as
the dilaton-axion with $\ell \rightarrow \ell+1$, while it does in the
extremal case.

\subsection{Scalar from two-form}

Substituting (\ref{tilde}) into (\ref{eqdil}) together with
(\ref{scalar}) yields
\begin{eqnarray}
\Big( \frac{1}{r^5} \partial_r fr^5 \partial_r +\omega^2 \frac{H}{f}
-\frac{\ell(\ell+4)}{r^2}+\frac{4}{r^2} (1-f) - \\ \nonumber
\frac{(H-1)[-f^2(15H+49)+f (\pm 16(\ell+2)H+30H+2)+H-1]}{4r^2 f H^2}
\Big) \tilde{\phi}=0. \label{genscalar}
\end{eqnarray}
We require four regions for matching, as defined in the previous section
for the two-form. In the outer region, the wave equation and solution are 
given by (\ref{outerV}) and (\ref{outersol}), respectively.
In the outer intermediate region, we may ignore the $\omega^2$ term in the 
wave equation:
\be
\Big( \frac{1}{r^5} \partial_r r^5 \partial_r
-\frac{\ell(\ell+4)}{r^2}-\frac{(H-1)[4\big( 1\pm (\ell+2) \big) 
H-12]}{r^2 H^2} \Big) \tilde{\phi}=0.
\ee
The corresponding wave function solution is
\be
\tilde{\phi}=C r^{\pm(\ell+2)-2} +D r^{\mp(\ell+2)-2} \big(1+
\frac{\ell+2}{\ell+2\pm 2} \frac{R^4}{r^4} \big).
\ee
Matching the outer and outer intermediate regions yields
\be
C=\frac{1}{(\ell+2)!}\big( \frac{\omega}{2}\big) ^{\ell+2},\ \ \ \ D=0,
\ee
for the positive sign and
\be
D=\frac{1}{(\ell+2)!}\big( \frac{\omega}{2}\big) ^{\ell+2},\ \ \ \ C=0,
\ee
for the negative sign.

Expressed in terms of the dimensionless quantity $z \equiv (r-r_o)/r_o$,
the wave equation in the inner intermediate region is
\be
\big( \frac{1}{z^3} \partial_z z^5 \partial_z +\frac{16\kappa^2}{z^2}
-\ell(\ell+4) \mp 4(\ell+2)-4 \big) \tilde{\phi}.
\ee
The corresponding wave function solution is
\be
\tilde{\phi}= A \frac{1}{r_o^2 z^2} H_{\ell+2\pm 2}^{(1)} \big( 
\frac{4\kappa}{z} \big),\label{intermscalar}
\ee
where we take the purely incoming solution. 

Matching the wave function across the two intermediate regions yields
\be
A_{\pm}=\frac{i \pi}{(\ell+2\pm 1)!(\ell+1\pm 1)!} \big( \frac{\omega
R}{2}\big)^{2\ell+4\pm 2}.
\ee
For $z \ll \kappa$, we expand the inner intermediate wave function
solution (\ref{intermscalar}):
\be
\tilde{\phi}= \frac{(-i)^{\ell+1/2} A}{r_o^2 \sqrt{2\pi \kappa}}
z^{-3/2} e^{4i\kappa /z} \big( 
1+\frac{i(\ell+3/2\pm
2)(\ell+5/2\pm 2)z}{8\kappa}+O(\kappa^{-2})\big). \label{smallscalar} \ee
For the inner region, we substitute (\ref{innersol}) into
(\ref{genscalar}) and express the result in terms of the dimensionless 
parameter $x$:
\begin{eqnarray}
x^2 f(x) \partial_x^2 \varphi+x\big[
5-(1-8i\kappa) x^{-4} \big] \partial_x \varphi +\Big(16 \kappa^2 h(x)
+ \\ \nonumber 4\big( 1-f(x)\big)-\ell(\ell+4) \mp 4(\ell+2)+\frac{15}{4}
f(x)-\frac{15}{2}-\frac{1}{f(x)}
\Big) \varphi=0. \label{innerequationscalar}
\end{eqnarray}
where $f(x)$ and $h(x)$ are given by (\ref{f}) and (\ref{h}), respectively.

Following the case for the dilaton, we plug the ansatz for the inner wave
function solution, given by (\ref{solform}), into
(\ref{innerequationscalar}) and solve for the amplitude $B$ by matching
the inner solution with the inner intermediate solution. The result is
that
\be
B=\frac{(-i)^{\ell+1/2}}{r_o^2 \sqrt{2\pi \kappa}} \big(
1-\frac{i(\ell+3/2\pm 2)(\ell+5/2\pm 2)}{8\kappa} \big).
\ee
Plugging amplitudes $A$ and $B$ into (\ref{prob}), we find that the
absorption probability is
\be
P^{near-extremal}= \Big( 1+\big( \frac{(\ell+3/2\pm 2)(\ell+5/2\pm 2)}{8
\kappa} \big)^2 \Big) P^{extremal}, 
\ee
where $P^{extremal}$ is the same as that given for the dilaton in
(\ref{extremal}) with $\ell \rightarrow \ell \pm 1$.
As with previous cases, as $\kappa \rightarrow \infty$, we
recover the result previously obtained for an extremal D3-brane.

We have shown that, for near-extremal D3-branes, the scalar from the
two-form does not have the same absorption probability as the
dilaton-axion with $\ell \rightarrow \ell \pm 1$, while it does in the
extremal case.

\section{Discussion}

We have obtained a simple relation between the extremal and near-extremal
absorption probabilities of a D3-brane:
\be
P^{near-extremal}= \Big( 1+\big( \frac{4\nu^2-1}{32 \kappa} \big)^2
\Big) P^{extremal}, \label{final}
\ee
where $\nu=\ell+2$ for the dilaton-axion, the vector from the two-form and
the antisymmetric tensor from the 4-form, $\nu=4$ for the two-form from
the antisymmetric tensor, and $\nu=\ell+4$ for the scalar from the
two-form. Note that (\ref{final}) has the same form as that of various
fields scattered by a four-dimensional $N=4$ supergravity equal-charge
black hole \cite{gubser2}.

As we would expect, a perturbation from extremality increases the
absorption probability. At near-extremality, the absorption probability of
the dilaton-axion, vector from the two-form and antisymmetric tensor from
the 4-form are identical, just as they are at extremality
\cite{poritz}. At extremality, the absorption probability of the scalar
from the two-form and the two-form from the antisymmetric tensor are
simply related to that of the dilaton-axion by a change in the partial
wave number $\ell \rightarrow \ell \pm 1$. For near-extremality, this
relation breaks down, and the scalar and two-form fields are absorbed more
than if this were the case.

We have found numerically that, further away from extremality, the
dilaton-axion, vector from the two-form and antisymmetric tensor from the
4-form no longer share the same absorption probability \cite{unpublished}. 
As can be seen from the effective potential (\ref{general}), this is due to 
a parameter $a$ which we conjecture depends on the angular momentum
polarization. $a=0$ for the dilaton-axion, $a=2$ for the tangential
components of the vector from the two-form, $a=4$ for the antisymmetric
tensor from the 4-form, and $a=6$ for the radial component of the vector
from the two-form. There are additional potential terms for the scalar
from the two-form and the two-form from the antisymmetric tensor. However,
these terms vanish for a chargeless D3-brane, enabling the scalar and
two-form fields to fit into the above scheme with $a=4$. It is rather
curious how the grouping of absorption probabilities changes between the
cases of extremal and chargeless D3-branes. At first look, it appears that
the time-component of the vector from the two-form does not fit into this
scheme, though certain potential terms may merely be equivalent to a
change in the partial wave number. We conjecture that the parameter $a$,
which only plays a role away from extremality, depends on the polarization
of the angular momentum. 

The absorption probability can be expressed in terms of Gamma functions:
\be
P \approx \frac{8 \pi^3 \kappa}{\big( \nu!(\nu-1)! \big)^2} \frac{|
\Gamma (\nu /2+1/4+2i \kappa )\Gamma (\nu/2+3/4+2i \kappa)|^2}{| \Gamma
(1+4i \kappa)|^2} \big( \frac{\omega r_o}{2} \big)^{2\nu}, \label{form}
\ee
so long as one keeps in mind that the formula holds only to order
$\kappa^{-2}$. For $\nu=\ell+2$, this absorption probability is the same as 
that derived in \cite{siopsis,siopsis2} for a minimally-coupled scalar. 
Siopsis expressed the absorption probability in the form (\ref{form}) in
order to interpret factors in terms of left and right-moving temperatures
$T_L$ and $T_R$. In particular, the general form of a black hole grey-body
factor is
\be
P_{b.h.} \sim \Big| \frac{\Gamma \big(
\frac{\ell+2}{2}+i\frac{\omega}{4\pi
T_L}\big) \Gamma \big( \frac{\ell+2}{2}+i\frac{\omega}{4\pi
T_R}\big) }{\Gamma \big( 1+\frac{\omega}{2\pi T_H}\big) }
\Big|^2. \label{bh}
\ee
Thus, by comparing (\ref{form}) with (\ref{bh}), we are led to the
conclusion that both left and right-moving modes contribute to the
absorption probability at temperatures
\be
T_L=T_R=T_H
\ee

\section{Acknowledgements}

We would like to thank M. Cveti\v{c} and H. L\"{u} for useful discussions,
and G. Siopsis for useful correspondence.

\end{document}